# Physiological Signal Processing in Heart Rate Variability Measurement: A Focus on Spectral Analysis

Amin Gasmi

1er Août 2022

## **Physiological Signals**

Feelings and emotions are significant determinants of the behavior of an individual. Human emotions can be established and recognized through several approaches, such as facial images, gesture, neuroimaging methods, and psychological signals [1]. The human body produces many physiological signals. These physiological signals can be categorized into eight major groups; the flow of body fluids, pressure, dimensions (e.g., imaging), bio-potential, temperature, impendence, and chemical concentration and composition. Due to the importance of these signals in the human body, it is of great significance that these signals be accessible for monitoring [2].

The accessibility and monitoring of these physiological signals are essential because; these signals may bear their origin the body parts (for example, infrared radiation), they may be internal (for example, those due to the change in pressure of the blood), and the signals may be from body tissue samples such as tissue biopsy or the blood. During the activities of various body physiological systems, the body releases physiological signals [3]. The body physiological systems' function is associated with information transfer to and from the respective system. Therefore, the information contained in the human physiological signals can be extracted and analyzed using various methods to establish the conditions and the working of such systems [4].

Extracting the information from these signal is necessary since it will aid in the understanding of the state and functioning of the system. The process through which this information is extracted varies depending on the energy content of the physiological system. Some are simple and easy, for example, torching the heart to feel its pulse, and to find and understand the heart beats' state [5]. However, others may be complex and therefore, will require sophisticated machines, especially those which involve the analysis of the tissue structure.

## **Classification of Physiological Signals**

Human physiological signals can be classified into seven categories depending on the type of energy. These classes are discussed as follows;

1. Biochemical Signals: - in this type of signal, the information from the signal is obtained either by analysis of the samples which are obtained from the body tissues or by chemical measurements done on the living tissues of the organism. Biochemical Signals is used in the determination of the concentrations of individual constituents in the blood tissues [6]. It is also used in measuring partial pressure carbon dioxide and oxygen in the respiratory system.
2. Bio-Optical Signals: - Bio-optical signals are usually produced in the body due to optical variations during physiological processes. An example of the application of these signals is in the measurement of the blood oxygenation [5]. Blood oxygenation is measured by comparing the measured quantity of diffused light and that which is reflected from the respective blood vessel.

3. Biomechanical Signals: - biomechanical signals are usually generated by the various mechanical functioning of a physiological system. According to [4] biomechanical signals are associated with motion, pressure changes or variations, displacements, and the flow of a physiological system. Analyzing the respiratory physiological system by considering the motion of the chest is an example of a real-life application of Biomechanical Signals.
4. Bioelectrical Signals: - these are the most known signals in the human body. Bioelectrical Signals are generated by muscle cells and nerve cells [1]. When cells undergo a state change, that is, from a rest to action under some conditions, bioelectrical signals generated. A fluctuating electrical field is usually generated when cells' potential changes, in the process of change of potential, bioelectrical signals are emitted. Example of Bioelectrical Signals is the electroencephalogram (EEG) and electrocardiogram (ECG) which are obtained from the bio signals of the brain and heart respectively [7].
5. Bio-magnetic Signals: - while the heart, lungs, and the brain functions, weak magnetic fields are usually generated [6]. Example of the biomagnetic signal is the magnetoencephalography which originates from the bio-magnetic signals of the brain.
6. Bio-impedance Signal: - Bio-impedance Signal arises from the voltage drop by the impedance of the tissue. For instance, the impedance of the skin is dependent upon three factors, blood distribution in the skin, the composition of skin, and the volume of blood through the skin [8]. Therefore, the measurement of the impedance balance of the skin helps in determining the skin conditions and working of the other physiological systems since skin carries out thermoregulation, which influences the operation of other systems.
7. Bio-acoustic Signals: - Bio-acoustic Signals are generated in the body due to the processes of blood and air flow systems [1]. For instance, the chest cavity process of opening and closing during respiratory process, and the blood flow in the chambers of the heart which generate unique acoustic signals.

In the analysis of these physiological signals, we must use a transducer for those signal which is not electrical since the digital computers accept only electrical signals [2]. A transducer is a device which converts a given signal into the electrical equivalent.

## Signal Processing Overview

Processing of the signal is designed to the intent of the measurement program. Signal Processing must be done with prior knowledge of the nature and the characteristic of the signal, especially the distribution of energy with frequency [9]. Signal Processing has a vital part in understanding the behavior of various physical processes. There are three major objectives of signal processing; these are summarized as follows:

1. To modify the signal and transform in a way that makes it easier to understand and interpret.
2. To exclude and remove unwanted components of the signal (which distorts the signal) such as high frequency or low-frequency signals.
3. To perform various operations on the signal that result in particular quantities of interest, such as time averages, corrections, root-mean-square values, or power spectra.

For effective signal processing, the signal must first be prepared. Preparation of signal for processing involve activities such as signal sampling, signal filtering, linearization, signal correction, and spectral analysis.

### Signal sampling

Signal sampling refers to the process of translating some portion of an input signal into discrete electrical signals for processing, storage or display. In signal processing, sampling is used to reduce a continuous signal into a discrete time signal. Sampling can be done for functions varying in space or time, and similar results are obtained. In a practical case, sampling of the continuous signal is achieved using Analog to Digital converter (ADC). Since using Analog to Digital converters have physical limitations, the sample data is usually not free from distortion. Some of the distortions associated with sampling include Aliasing, Aperture error, Jitter, and noise. For accuracy to be maintained, this distortion must be removed. For instance, the noise in the sample is removed by use of signal filters. In physiological signals, there are three major types of sampling techniques, natural sampling, impulse sampling, and flat top sampling.

Transformation using Fourier method is used to help in understanding of sampling and re-sampling of discrete signals. From a continuous process, sampling creates discrete signals. This is achieved using two methods; down sampling (decimation) which is the process of sub-sampling signals of discrete nature or up-sampling which is the process in which zero is introduced between samples thereby making the signal to be longer [10]. In most signal processing, down-sampling is used and thus we consider the process f down sampling.

Down - Sample process

We consider down-sampling a signal I [n] of length N:

Our objectives is to reduce N by ns factor. We note that ns must be a divisor of N.

We define a comb function as:

$$C(n; n_s) = \sum_{m=0}^{(N/n_s)-1} \delta_{n, mn_s}$$

Steps

1. We introduce zeros in I[n] at the unwanted samples.

$$g_0[n] = C[n; n_s] I[n]$$

2. We down-sample signal $g_0$;

$$g[m] = g_0[mn_s] \text{ for } 0 \leq m < N/n_s$$

Applying frequency domain analysis (Fourier transform) to the two steps, we must define tow propositions as follows:

Proposition 1: Comb function's Fourier transform is also comb function:

$$F(C[n; n_s]) = \frac{N}{n_s} C[k; N/n_s]$$

Where k = wave number.

Proposition 2: Pointwise product and Fourier transforms convolution

Suppose f [n] and g [n] are two different signals having length N (which is N-periodic under extension). Then

$$F(f[n]g[n]) = \frac{1}{N} \times F(f) \times F(g)$$

$$\equiv \frac{1}{N} \sum_{j=-N/2}^{(N/2)-1} \hat{f}[j]\hat{g}[k-j]$$

Where $\hat{f}$ and $\hat{g}$ are the Fourier transforms of $f$ and $g$, respectively.

Applying Fourier analysis for step 1.

From step 1;

$$g_0[n] = C[n; n_s] I[n]$$

Therefore, by proposition 1 and 2 above,

$$F(g_0) = F(C[n; n_s] I[n])$$

$$= \frac{N}{n_s} \frac{1}{N} C[k; N/n_s] . \hat{I}[k]$$

$$\equiv \frac{1}{n_s} \sum_{j=N/2+1}^{(N/2)} C[j; N/n_s] \hat{I}[k-j]$$

$$= \frac{1}{n_s} \sum_{r=-r_0+1}^{n_s-r_0} \hat{I}[k - r\frac{N}{n_s}]$$

Where,

Since I[k], possesses periodicity, we can make use any integer $r_0$ $(e.g.\ r_0 = n_s/2\ for\ even\ ns)$. diagrammatically, this is illustrated below;

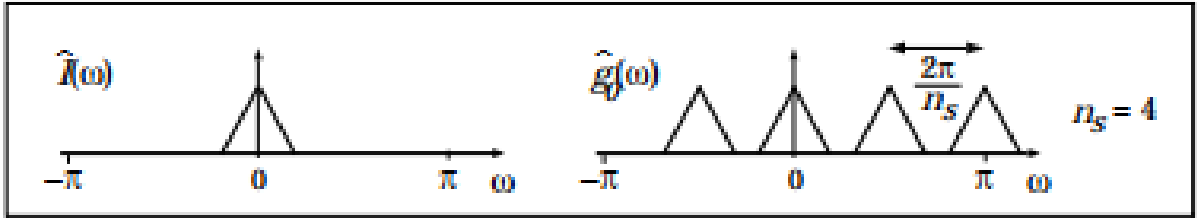

**Fig 1.** Fourier analysis of step 1. [10]

Therefore $g_0$ [k] is the summation of replicas I [k – rN/ns] of the FT of the signal originally present I, spaced by wave number $N = n_s$ or by frequency;

$$\omega_s = \frac{2\pi}{n_s}.$$

It should be noted that the Fourier transform $g^0$ possesses 2π period, thus the sketch for contribution above for ω greater or equal to π can easily be shifted 2π to the left.

When we apply Fourier transform in the second step, the samples from $g_0$ [n] is dropped. This is because it has been set to zero by the comb function C [n; $n_s$]. That is

$$g[m] = g_0[mn_s]\ for\ 0 \leq m < N/n_s$$

When using the Fourier transform, it is simple to show

$$\mathcal{F}(g)[k] = \hat{g}[k] = \hat{g}_0[k],\ \text{for}\ -N/(2n_s) \leq k < N/(2n_s).$$

This can be written to give the below diagram

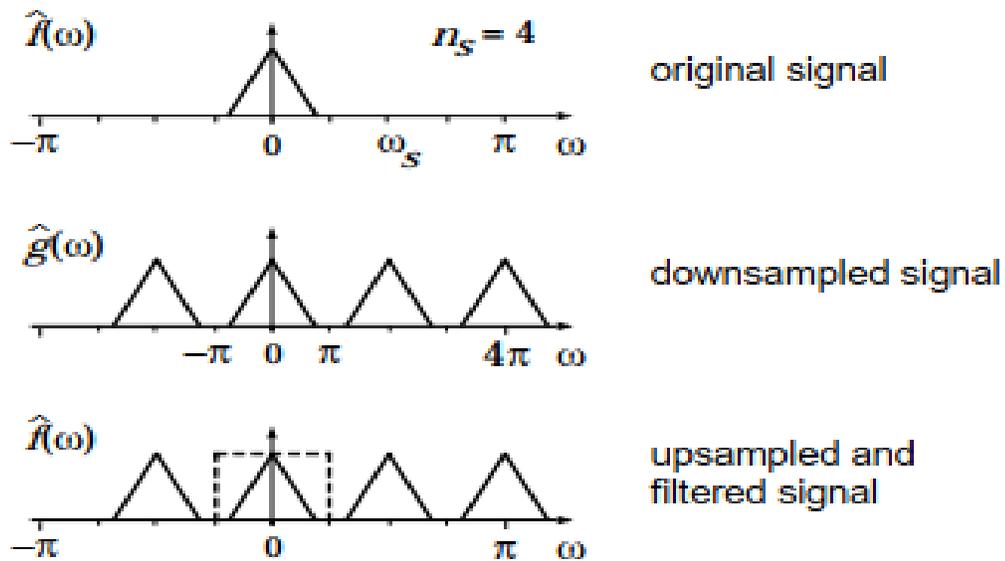

**Fig 2.** Fourier analysis of step 2. [10]

**Signal Filtration**

Filtering in signal processing is done primarily to remove the low-frequency components or high-frequency components [11]. These unwanted frequency components are usually called signal noise. In most cases, the noise is found not to be related to turbulent flow signal, and are filtered to exclude signals unrelated to a particular event in case of conditional sampling. In digital filtering processes, high-frequency noise is filtered out to pass without attenuation, while low-frequency noise is filtered by choice of sampling time or high pass filtering. Signal filtration can be achieved by the use of both analog filters and digital filters. However, digital filters are preferred to analog counterparts [12].

Digital filters form a very significant part of Digital Signal Processing (DSP). Their excellence performance explains the popularity of Digital Signal Processing. They have two primary functions, signal restoration, and signal separation. Signal separation is important, especially when a signal has been contaminated with various interference such as noise, or other unwanted signals. For instance, in the physiological process where a device is used to measure the electrical activity and the state of a baby's heart (EKG) while the baby is still in the womb, the raw signal is likely to be corrupted by the breathing and heartbeat of the mother. Overcome the signal from the mother calls for the use of a filter [13]. A filter might, therefore, be used in separating these signals so that the signals be analyzed individually.

Signal restoration is usually applied in cases where a signal has been distorted in some way. In physiology, for instance, measuring of the heart rate variability done using poor equipment may be filtered to using digital filters to better represent the variability as it actually occurs [14]. It is acceptable in Digital Signal Processing to say that a filter's input signals and output signals are in the time domain. This is because digital signals are usually created through sampling at regular time intervals. Implementing digital filters is usually an issue, especially when dealing with the human physiological process [15]. However, the most prominent and straightforward way of implementing a digital filter is by convolving the input signal with the digital filter's impulse response.

Fourier transform is important in signal filtering since the convolution in the time domain can be replaced with a multiplication in a frequency domain [16]. The dime domain filtering equation can be given as below;

$$y(t) = x(t) * h(t) \xleftrightarrow{\mathcal{F}} Y(\omega) = X(\omega) \cdot H(\omega)$$

The above time doain can be transform into frequency domain and thus can be analyzed using Fourier transform

$$y[n] = x[n] * h[n] \xleftrightarrow{\mathcal{F}} Y[k] = X[k]H[k]$$

**Signal Normalization**

Whenever there are many signals emanating from a source, it is important to compare such signals. Signal normalization is necessary when we are to compare it with respect to other signals. Normalization is simply the scaling of the signals in an identical level [9].

Normalization is basically bringing signals of different range to the same range or to a predefined range, in signal processing, normalizations eases and robustifies the comparison process [10]. Normalization also aid visual and analysis.

Given two signals below,

$$y_1(t) = x_1(t) + n_1(t)$$

$$y_2(t) = x_2(t) + n_2(t)$$

Where;

$n_1(t) \sim N(0, \sigma_1^2)$, $n_2(t) \sim N(0, \sigma_2^2)$, and $x_1(t)$ and $x_2(t)$ are deterministic.

In order to compare the signal, to obtain a normalization, we compute;

$$z_1(t) = \frac{y_1(t)}{\sigma_1}$$

$$z_2(t) = \frac{y_2(t)}{\sigma_2}$$

Now $z_1(t)$ and $z_2(t)$ are observed to have a noise with variance 1. This method has an advantage that if $z_1(t)$ is larger than $z_2(t)$, then $z_1(t)$ has a better signal to noise ratio (SNR).

**Signal Linearization**

Linearization is done primarily to linearize the non-linear relationship that may be existing between the variables. During linearization, the fluctuating electrical signals are converted into one in which the voltage is proportional to the velocity, i.e., linearize the signal. Linearization of the signals can be achieved through digital linearization or analog linearization.

Many transducers used in the measurement of physiological signals have an output that is a non-linear function of the measured quantity input [17]. The non-linear signal can be linearized using a special operational amplifier. The linearizing amplifier should have a configuration that have an equal but opposite non-linear relationship between the amplifier input and output terminals.

We consider a signal which has an exponential relationship between the output signal and the input signal. The relationship can be mathematically related as

$$V_0 = Ke^{-\alpha Q}$$

Where Q and V0 are the signal input and output respectively, K and α are constants. If a diode is place between the input and the output terminal of the operational amplifier as shown below, the relationship between the amplifier output voltage V0 and input voltage V1 can therefore be expressed as;

$$V_0 = C \log_e V_1$$

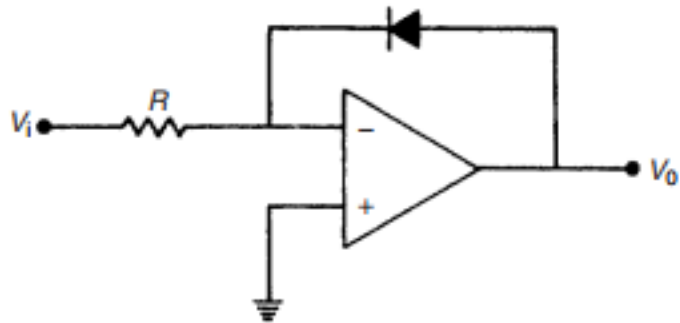

**Fig 3.** Operational amplifier connected for signal linearization [18].

If the output of the transducer with characteristic given in the first equation above is conditioned by an operational amplifier of characteristic given by the second equation above, the output V0 will be given by the equation,

$$V_0 = C \log_e K - \alpha C Q$$

The equation shows that now the output signal v0 now varies linearly with the input Q but with an offset of C loge K. this indicates that linearization of signals is usually achieved by the use of operational amplifiers [18].

**Spectral Analysis**

The spectral analysis involves the analysis which is carried out in terms of the spectrum of frequencies or other related quantities like energy. The main function of spectral analysis is to decompose a time series into periodic components, which will be easy to analyze [19]. Decomposition can be achieved through various ways, but the most common method of decomposition of times series is through a regression, where the time series is regressed on a set of cosine and sine waves.

A discrete or continuous time – series of the form x = x (t) or generally $x_n$ = {$x_0$, $x_1$,. . .}, can be analyzed while utilizing both frequency-domain descriptions and time-domain descriptions. The analysis, which involves frequency-domain descriptions, is referred to as spectral analysis. Spectral analysis reveals some features of a time-series which are difficult to be noted when the analysis is carried out using time-domain description. Spectral analysis is an important tool as it is used to solve a variety of practical problems, especially in science and engineering. Examples of areas where Spectral analysis is used include but not limited to the study of vibrations, interfacial waves, physiological processes of the heart and brain (e.g., HRV), and stability analysis.

In practical cases of spectral analysis, a time-series is disintegrated into sine and cosine wave components. This is achieved by getting the summation of all the weighted sinusoidal functions. These sinusoidal functions are called spectral components. Either a density of spectral components or spectral density function forms the weighting function in such disintegration. The use of Fourier transform has been the actual and most practical method for decomposition of time-series into a sum of weighted sinusoidal functions. The advantage of using Fourier transform is that it has both discrete and continuous versions which can easily correspond to discrete time-series of type:

$$x_n = \{x_0, x_1, \ldots\},$$

And continuous time-series of type $x = x(t)$ respectively [20]. Many engineering data analysis is currently done using dedicated computers which only accept discrete data type, and therefore, we apply discrete Fourier transform (DFT) whose spectral analysis is discussed as follows.

**Fourier Transform**

Assuming we have a discrete time – series with N samples and $T_s$ as sampling time interval between any two successive samples. This can be represented as;

$$x_n = \{x_0, x_1, \ldots, x_{N-1}\}$$

Each Fourier transform has its inverse and therefore, according Fourier analysis' mathematical theory, the time-series above can be written in form of finite inverse Fourier transform below;

$$x_n - \bar{x} = \frac{1}{N}\sum_{m=1}^{N-1} G(m) e^{j2\pi mn/N} \quad (n=0,1,\ldots,N-1)$$

Where;

X bar = the mean value of the time-series

$j = \sqrt{-1}$ is the symbol of a complex number

$e^{j\theta}$ = a complex sinusoidal function i.e. $e^{j\theta} = \cos\theta + j\sin\theta$

G (m) = weighting function. it can also be termed the spectral density function.

The spectral density function is calculated using the finite Fourier transform equation below;

$$G(m) = \sum_{n=0}^{N-1}(x_n - \bar{x})e^{-j2\pi(m/NT_s)nT_s} = \sum_{n=0}^{N-1}(x_n - \bar{x})e^{-j2\pi mn/N}$$

m = 0, 1, ...., N-1;

From the equation:

$m/NT_s = f_m$ represent the discrete frequency.

$nT_s = t_n$ represent discrete time.

Again we must note that $(x_n - x)$ which appears in the first equation and G (m) which is in the second equation forms equal measures in the time domain and frequency domain respectively. $(x_n - x)$ and G(m) are related by the Fourier transform.

This operation finds its applications in various areas; however, its direct computation based on its definition is usually so slow and therefore impractical. The Fast Fourier Transform (FFT) is usually employed as a way to compute the DFT more quickly since it reduces the computation time and cost.

**Frequency Spectrum**

From the above definition, the Fourier transform G(m) is a complex-valued function, and a frequency spectrum is obtained when we plot G(m) against $f_m$. the polar form of G(m) is written as;

$$G(m) = |G(m)| e^{j\angle G(m)}$$

Where, G(m) and magnitude |G(m)| are the Fourier transform's phase angle and spectral modulus respectively. When we plot |G(m)| against $f_m$ the resulting plot is termed the amplitude or the magnitude spectrum, and when we plot G(m) against $f_m$, the resulting plot is termed the phase spectrum.

## Power Spectrum

Power spectrum $R_{xx}(k)$ of a time series $x(t)$ is a description of power into frequency components making up the signal. Applying Fourier analysis, it is clear that any physical signal can be decomposed into a number of discrete frequencies (usually termed spectrum of frequencies) over a continuous range.

The definition of the discrete time-series auto-correlation function given in the first equation is defined as

$$R_{xx}(k) = \sum_{n=0}^{N-1-k} (x_n - \bar{x})(x_{n+k} - \bar{x}) \quad (k = 0, 1, \ldots, N-1)$$

The auto-correlation function has its Fourier transform defined as;

$$R_{xx}(m) = \sum_{k=0}^{N-1} R_{xx}(k) e^{-j2\pi mk/N} \quad (m = 0, 1, \ldots, N-1)$$

Where $R_{xx}(m)$ is the function of the power spectral density. When we plot $R_{xx}(m)$ against $f_m$, the resulting plot is the power spectrum. This power spectrum corresponds to the time-series formula, which is elaborated in the first equation above [5]. A close consideration of both power spectrum $R_{xx}(m)$ and frequency spectrum $G(m)$ reveals a mathematical relation between the two spectra. The relationship is given by the below equation.

$$P_{xx}(m) = |G(m)|^2 \quad (m = 0, 1, \ldots, N-1)$$

From the equation, it can be inferred that $P_{xx}(m)$ is a function which is real-valued. The function's phase is zero. Therefore, we can conclude that the average measure frequency-domain features of a given time-series forms the power spectrum.

## Cross Spectrum

Cross Spectrum correlate two sets of time - series data. For any two time-series data set:

$$x_n = \{x_0, x_1, ..., x_{N-1}\} \text{ and } y_n = \{y_0, y_1, ..., y_{N-1}\}$$

The cross-correlation function is expressed as;

$$R_{xy}(k) = \sum_{n=0}^{N-1-k} (x_n - \bar{x})(y_{n+k} - \bar{y}) \quad (k = 0, 1, ..., N-1)$$

And its Fourier Transform is given by

$$P_{xy}(m) = \sum_{k=0}^{N-1} R_{xx}(k) e^{-j2\pi mk/N} \quad (m = 0, 1, ..., N-1)$$

Where $P_{xy}(m)$ is the cross-spectral density function or simply the cross spectrum. Cross Spectrum is usually a complex-valued function. The common frequencies observed in both the time-series $x_n$ and $y_n$ are usually represented by the cross spectrum.

## Utility of Using FFT in Physiological Signal Processing

The Fourier transform is the most commonly used transform in the frequency domain analysis of a physiological signal. Sinusoidal functions form the basis functions of the Fourier transform, and therefore, it is frequently used for the study of Physiological Signal [21]. In utilizing the Fast Fourier Transform, projections are computed for the given signal function represented as a function of time $x(t)$ onto the complex exponential basis function of frequency $\omega$. When establishing an understanding of a system, it is essential to consider the variation in frequency over the duration of a signal. The waveform the usually characterize a Physiological signal such as Radial Pulse or signal may be affected by many factors such as the contacting force between the sensor and the measurement site.

In research done by [21], the pulse signal was recorded with an advanced pulse detection system for different contact pressures on the radial artery of the healthy subjects. In the research, it was found out that as the contact pressure increases on the wrist, the amplitude of pulse increases at first, it then becomes maximum at a specific pressure and then the amplitude starts to decrease especially when the contact pressure starts to increase during a physiological process. Fast Fourier transform is used in the physiological process in analyzing pulse, which shows the variation in fundamental frequency and other harmonic frequencies when various pressure quantities are applied on the radial pulse.

Fast Fourier Transform is used in various physiological signal processing, such as in the analysis of the rhythmic contraction and relaxation of the heart, which causes the travelling pressure wave in the heart. In many physiological signal analysis, the transducer is positioned on the skin where it detect the signal of physiological systems like respiration, heartbeat rate, and heart rate variation [1]. The common feature in these physiological process is that they produce pulse wave. Transducers can be position and adjusted to captures an appropriate signal. The Fast Fourier Transform (FFT) can be used to obtain and determine the spectrum of these pulse waves in any physiological process.

Fast Fourier transform (FFT) is commonly used to analyze the spectral content of any deterministic bio-signal (signal contaminated or not contaminated by noise). Many signals are, in most, cases analyzed using Discrete Fourier transform (DFT) [22]. However, DFT is too costly and time-consuming method; thus, we apply a Fast Fourier transform, which will help in reducing the time and cost of analyzing such signals. Discrete Fourier transform allows for the decomposition of discrete time signals into sinusoidal (sine or cosine wave) components [23]. The frequencies of these sinusoidal components are usually multiples of a fundamental frequency of the signal.

The phases and amplitudes of such sinusoidal components are estimated using the Discrete Fourier Transform and are mathematically represented as;

$$X(k) = \frac{1}{N} \sum_{n=0}^{N-1} x(n) e^{-j(2\pi kn/NT)}$$

For a given physiological (bio-signal) represented as $x(n)$, having a sampling period $T$ with $N$ number of total samples, it, therefore, implies that $NT$ is the total duration of the signal segment. The signal spectrum $X(k)$ can be estimated at multiples of $fs/N$, where $fs$ represents the sampling frequency. When applying Fast Fourier transforms in the computation of Discrete Fourier transform, it is vital to appreciate the limitation in resolution (for the length of the signal) and the signal windowing effects on the exactness of the projected spectrum [22]. This is because Fast Fourier Transform does not work so well with short – duration signals.

In physiological signal processing, direct application of Fast Fourier Transform (FFT) on the signal implicitly assumes that a rectangular window, with a unity value over the duration of the signal and zero value at all other times, multiplies the signal. However, a multiplication in the time domain will result in a convolution operation in the frequency domain between the spectrum of the rectangular window and the original spectrum of the signal $x(n)$, their care must be taken during such analysis. Since the spectrum of the rectangular window is usually a Sine function which consists of decaying sinusoidal ripples, the resulting convolved spectrum $X(f)$ can, therefore, be a distorted version of the original spectrum of $x(n)$. Due to the convolution operation, the content of the spectrum from one frequency component (usually the dominating spectral peak) specifically, tend to leak into neighboring frequency components [19]. Because of such leakage, it is often advisable to window the signals, especially when a dominating spectral peak is expected to be positioned adjacent to one with a lower amplitude in the signal spectrum.

# Application FFT in heart rate variability

## Heart Rate Variability

Human physiological (biological) systems function in such a way that their complexity requires mathematical analysis. The functioning of the brain, heart and other parts are so complex to be easily comprehended [24]. Under conditions of rest or work, the temporal distances of successive heartbeats are subject to fluctuations, thereby forming the basis of Heart Rate Variability (HRV). In normal conditions, the human is persistently exposed to highly changing and dynamic situational demands [20]. With these demands in mind, HRV can, therefore, be considered as the human organism's ability to cope with and adapt to continuous situational requirements, both physiologically and emotionally. Generally, individuals who have highly variable heartbeat rates are considered as healthy organisms and thus are well able to cope up with different situational demands, however, according to [24] high HRV is not always better as pathological conditions can also produce high HRV. On the contrary, individuals with low HRV are considered to be susceptible to health-related complications.

Heart rate variability is usually confused with heart rate. However, there is a big difference on the two, and both metrics have pros and cons which should be put in mind before deciding which metric to be used when analyzing the heart performance [25]. While heart rate (HR) measures heart beats per minute, heart rate variability (HRV) is the measure of the specific changes in time (variability) between successive heartbeats. Heart Rate Variability, due to its complexity in measurement and focus on imperceptible changes that occurs between each heartbeat, it requires higher degrees of accuracy compare to the heart rate measurements. HRV should be applied at rest and can be used in determining the heart's readiness to tolerate stress on a given day [26]. With this in mind, HRV is commonly applied in the optimization and individualize training programs which are based on an individual's readiness or recovery state.

Heart Rate Variability (HRV) indexes neuro-cardiac function and the variability is generated by the interaction of the heart and the brain, and dynamic non-linear autonomic nervous system (ANS) processes. From this interaction, HRV can further be looked at as a developing feature of mutually dependent regulatory systems whose operation is on a different time scales to help an individual to adapt to the varying psychological and environmental challenges. Heart Rate Variability reflects regulation of autonomic balance, gas exchange, blood pressure (BP), heart, gut, and vascular tone, which is dependent to the diameter of the blood vessels [27]. The Vascular tone is responsible for regulating Blood pressure, and possibly facial muscles [24].

**Application of FFT in HRV Analysis**

Fast Fourier Transform (FFT) relies on the frequency-domain HRV analysis techniques. It requires re-sampling of the inherently unevenly sampled heartbeat time-series (RR tachogram) to produce an evenly sampled time series of the heartbeat. However, re-sampling of the heartbeat time – series is found to produce a substantial error when estimating an artificial RR tachogram.

In the analysis of heart rate variability HRV using the frequency domain, there are two main methods usually used; the Fourier techniques (e.g. Discrete Fourier Transform – DFT and Fast Fourier Transform - FFT), and the auto-regressive (AR) spectral estimation. However, no official publication from the authoritative bodies have recommended the use of one method over the other and thus these two methods are equally used in the Heart Rate Variation analysis. In a research conducted by [28] showed that FFT and AR methods could provide a comparable measure of the Low Frequency and High-Frequency metrics on linearly resampled 5 minute HR segments across a patient population with a wide variety of ages and medical conditions. This paper focusses on the use of Fast Fourier transform on the analysis of Heart Rate Variation.

Just like any other physiological signal processing, Heart Rate Variability (HRV) can be analyzed using both frequency & time domain analyses. Time domain index for Heart Rate Variability primarily quantifies the amount of variability in measurements of the inter-beat interval (IBI) of the heart, which can be described as the periodic time between successive heartbeats. The resulting values from such measurement are expressed in original units.

However, if the distribution is abnormal, the natural logarithm (ln) of original units can be used to express the values thereby achieving a more normal distribution.

Frequency domain measurement, on the other hand, estimates the distribution of absolute and relative power into four frequency bands of the heart rate (HR) oscillations. The four bands comprise Ultra Low Frequency (ULF) band, Very Low Frequency (VLF) band, Low Frequency (LF) band, and High Frequency (HF) band. These bands vary from each other in terms of their operational frequency and the methods of capturing the band. For instance, the Ultra-Low Frequency (ULF) band (≤0.003 Hz) is used to index the variations in IBIs with a period from approximately five minutes to twenty-four hours. We shall consider some of these bands

*Very Low Frequency (VLF) band:* - the frequency range for this band is (0.0033 – 0.04 Hz). The band requires a five minutes recording period at the minimum level. However, over 24 hours is the best way to monitor the band. For LVF band, there exist about 0–12 complete periods of oscillation within a 5-min sample. Very Low-Frequency power is more associated with all-cause mortality than Low Frequency or High-Frequency power. While implementing a Fast Fourier Transform (FFT) in Very Low-Frequency band, it must be noted that FFT works on a certain number of data points at a time. For FFT, the number of data points is directly proportional to the frequency resolution. Small FFT sizes do produce a higher amplitude accuracy with low-frequency resolution. The bin width (sampling rate divided by FFT size) is determined by the FFT size. For example, with FFT size of 1K (1024), and a sampling rate of 200/s, bin width is found to be 0.2 Hz, which implies that frequencies below 0.2 Hz cannot be seen.

*Low Frequency (LF) band:* - the frequency range for this band is (0.04 – 0.15 Hz). A two minutes a recording period at the minimum level is required for the band. The region under which this band falls was originally known as the baroreceptor range since it majorly reflects baroreceptor activity during resting conditions. In pomological processes, low-frequency power is mainly produced by the SNS, PNS and BP regulation *via* baroreceptors. However, above 0.1 Hz, the SNS seems not to produce rhythms. The parasympathetic system, on the other hand, is observed to be able to affect heart rhythms down to 0.05 Hz. A practical consideration in the physiological process during respiration where at low respiration rates, the nerve functions can produce oscillations easily in the heart rhythms that cross over into the LF band.

*High Frequency (HF) band:* - this band is termed the respiratory band since it matches the heart rate variations related to the respiratory cycle. The frequency range for this band is (0.15–0.40 Hz). The band requires a recording period of 1 min at the minimum level. Infants and young children do breathe faster compared to the adults, the resting range for the band can be adjusted to 0.24–1.04 Hz. The acceleration and deceleration of the Heart rate during inspiration and expiration are due to the introduction and removal of oxygen in the lungs respectively. As the inhalation process occurs, the cardiovascular center inhibits vagal outflow resulting in the speed up of heart rate [29]

Conversely, when one breathes out, the cardiovascular center restores vagal outflow resulting in slow down of the heart rate by releasing acetylcholine. High-Frequency oscillations can be virtually eliminated by total vagal blockage and therefore reduces power in the Low-Frequency range. High-Frequency band power may rise during the night and fall during the day. Lower high frequency power can be correlated with anxiety, panic, stress, and/or worry [23].

*Total spectral power:* - spectral power is the energy of the signal found within the frequency band. Frequency domain measurements are expressed both as absolute power or relative power. The total spectral power is therefore the summation of all the energy in the Ultra-Low, Very Low, Low, and High-Frequency bands for 24 h and the Very Low, Low, and High-Frequency bands for short-term recordings [20]. Fast Fourier transform (FFT) is applied in the total spectral power through its application in the individual bands. Thus after FFT is applied in the individual frequency band, in the total spectral power, we sum the frequency band [30]. Therefore, FFT is not directly applied in the total spectral power but indirectly through the individual frequency bands.

*Low Frequency – High-Frequency Ratio (LF/HF ratio):* - according to [31], the original intention of obtaining the LF/HF ratio was for the estimation of the sympathetic nervous system (SNS) and peripheral nervous system (PNS). LF/HF ratio was originally obtained on the basis of a twenty-four-hour recording. During this period, LF power is contributed by SNS and PNS activities, while HF power is mainly contributed by PNS activity.

The Low Frequency to High Frequency ratio is underlying in the assumption that Low-frequency power may be can be produced by SNS while High-frequency power is produced by PNS. Using the assumption, a low LF/HF ratio implies parasympathetic dominance [31]. This, in the practical case, is experienced when individuals conserve energy and engage in tend-and-befriend behaviors. On the other hand, a high LF/HF ratio implies the presence of sympathetic dominance, a practical case in this scenario is when an individual engages in parasympathetic withdrawal [24].